\documentclass {amsart}
\usepackage[pdftex]{graphicx}

\newenvironment{alignedt}{
                \setlength{\arraycolsep}{0.1em}
                \begin{array}[t]{rll}
                \displaystyle
        }{
                \end{array}
        }

\newcommand{\sgn}{\text{sgn}}
\newcommand{\median}{\text{median}}
\newcommand{\pz}{\phantom{-}}
\newcommand{\pb}{\phantom{\;\;}}
\newcommand{\mubar}{\bar{\mu}}
\newcommand{\nubar}{\bar{\nu}}
\newcommand{\muhat}{\widehat{\mu}}
\newcommand{\nuhat}{\widehat{\nu}}
\newcommand{\var}{\text{var}}
\newcommand{\Ij}{\mathcal{I}(j)}
\newcommand{\Ji}{\mathcal{J}(i)}
\newcommand{\Jip}{\mathcal{J}(i')}

\makeatletter
\def\Ddots{\mathinner{\mkern1mu\raise\p@
\vbox{\kern7\p@\hbox{.}}\mkern2mu
\raise4\p@\hbox{.}\mkern2mu\raise7\p@\hbox{.}\mkern1mu}}
\makeatother

\title{A Regression Approach to Fairer Grading}
\author{ Robert J. Vanderbei
\and
	Gordon Scharf
\and
	Daniel Marlow
}
	\thanks{Department of Operations Research and Financial Engineering,
	Princeton University,
	Princeton, NJ 08544 ({\tt rvdb@princeton.edu})
	}
	\thanks{\tt gordon.scharf@gmail.com}
	\thanks{Department of Physics,
	Princeton University,
	Princeton, NJ 08544 ({\tt marlow@princeton.edu}).
	}

\begin{document}

\maketitle

\begin{abstract}
In this paper we describe a statistical procedure to account for differences in
grading practices from one course to another.  The goal is to define a course
``inflatedness'' and a student ``aptitude'' that best captures ones intuitive
notions of these concepts.
\end{abstract}



%
%
%
%

\section{Introduction}

Course assessment and grading policy are topics of great interest to most
students.
Mathematical models that address inherent unfairness in the assessment
process provide an excellent example of regression that can be taught in
undergraduate statistics and/or optimization courses.
In fact, one of us (Scharf) was a junior contemplating what would make an
interesting senior thesis and after a casual dinner conversation with classmates 
came up with the idea that a statistical method to adjust
student grade-point averages according to the difficulty of the courses taken
could lead to a very interesting thesis.  This article, while it highlights a
different statistical approach than the one originally proposed, is an outgrowth of that
thesis.  

Suppose a student takes both course X and course Y and gets a higher grade in course X
than in course Y.   Based on just one student, it is likely that the
student simply has more aptitude for the material in course X than for the material in
course Y.   But, if most students who took both courses X and Y got a better
grade in course X than in course Y, then one begins to think that course X
simply employed a more inflated grading scheme.

Consider for example, a school with only four students: John, Paul, George, and
Ringo.  Suppose that this school only offers six different courses from
which the students select four to take.  The students made their selections,
took the courses, and we now have grading information as shown in Table
\ref{tab1}.
From this table, we see that George and Paul have received the same grades (in
different courses) and so their {\em grade-point averages} (GPA's) 
are the same.
Furthermore, John's grades are only slightly better and Ringo's grades only
slightly worse than average.   But, it is also clear that the Math class gave
lower grades than the Economics course.   In fact, there is a linear progression
in grade-inflation as one progresses from left to right across the table.
Taking this into account, it would seem that John took ``harder'' courses than
Paul (the quotes are to emphasize that a course that gives lower grades is not
necessarily more difficult even though we shall use such
language throughout this paper), who took harder courses than George, who took
harder courses than Ringo.   Hence, GPA does not tell an unbiased story.   John
did the best in all of his courses, in many cases by a wide margin.   Ringo, on
the other hand, did the worst in all of his classes, again by a wide margin.
It is clear that John is a much better student than Ringo---better to a degree
that is not reflected in their GPA's.   
\begin{table}
\begin{center}
\begin{tabular}{l|llllll}
       & MAT   & CHE   & ANT   & REL   & POL   & ECO  \\[0.1in] \hline
John  & B$-$  & B     & B$+$  & A$-$  &       &      \\ 
Paul  & C$+$  & B$-$  &       & B$+$  & A$-$  &      \\
George&       & C$+$  & B$-$  &       & B$+$  & A$-$ \\
Ringo &       &       & C$+$  & B$-$  & B     & B$+$ \\
\hline
\end{tabular}
\end{center}
\vspace*{0.2in}
\caption{Grading data from Beatle University.  The six courses are Math
(MAT), Chemical Engineering (CHE), Anthropology (ANT), Religion (REL), Politics
(POL), and Economics (ECO).}
\label{tab1}
\end{table}

Our aim is to develop a model that can be
used to infer automatically the sort of conclusions that we have just drawn for
this small example.  
Of course, one must consider the simplest suggestion of just computing averages
within each course.  Clearly, in Table \ref{tab1}, the Math course gave grades a full
letter grade lower than the Econ course.  One could argue that that is all one
needs---just correct using average grades within each course.   But, one can
easily modify the simple example shown in Table \ref{tab1} to make all the courses have
the same average grade and all of the students have the same GPA but for which
there is an obvious trend in the true aptitude of the students.  Table
\ref{tab2} shows
one rather contrived way to do this (using an unbounded list of courses and
students).

Finally, the model must be computationally tractable so that it can
be run for a school with thousands of students taking dozens of courses (over
four years) selected from a catalogue of hundreds of courses.
\begin{table}
\begin{center}
\begin{tabular}{l|lllllllllll}
       & $\cdots$ & MAT   & CHE   & ANT   & REL   & POL   & ECO  & HIS & $\cdots$ \\[0.1in] \hline
       & & & & & & & & \phantom{MAT} \\
$\vdots$  & $\ddots$ & $\ddots$  & $\ddots$  & $\ddots$  &       &       &      \\ 
John  & & B$-$  & B     & B$+$  & A$-$  &       &      \\ 
Paul  & &       & B$-$  & B     & B$+$  & A$-$  &      \\
George& &       &       & B$-$  & B     & B$+$  & A$-$ \\
Ringo & &       &       &       & B$-$  & B     & B$+$  & A$-$  &       &      \\
$\vdots$  & &       &       & & & $\ddots$ & $\ddots$  & $\ddots$  & $\ddots$  \\ 
\hline
\end{tabular}
\end{center}
\vspace*{0.2in}
\caption{A school with an infinite number of students and an infinite selection
	of courses.  Every student has the same GPA and every course has the
		same course average.  Yet, John is smarter than Paul is
		smarter than George is smarter than Ringo and Math is harder than
		Chemical Engineering is harder than Anthropology etc.}
\label{tab2}
\end{table}

\section{The Model}

We assume that there are $m$ students and $n$ courses.  The data consists of the 
grades for all courses taught.  For each course, we assume that we have grading
data for every student who took that course.  But, we do not assume that every
student takes every course offered.  In fact, we assume quite the opposite,
namely, that each student only takes a small sample of the complete
suite of courses offered.  

We assume that each student has an aptitude\footnote{Several colleagues have
pointed out the obvious fact that aptitude varies from subject to
subject.  We are not trying to capture this variation.  In this
paper, we consider ``aptitude'' to be a synonym for ``modified
GPA''---a one-dimensional parameter that could be used to determine class rank,
awards, etc.}
$\mu_i$, $i=1,2,\ldots,m$, which is
unknown to us and which we wish to estimate, and that each course has an
inflatedness $\nu_j$, $j=1,2,\ldots,n$, which is also unknown to us and also of
interest to estimate.  We assume that each grade $X_{ij}$ can be approximated as the sum
of the student's aptitude plus the course's inflatedness:
\begin{equation}
    X_{ij} = \mu_i + \nu_j + \epsilon_{ij}, \qquad (i,j) \in \mathcal{G}
\end{equation}
where $\mathcal{G}$ represents the set of student-course pairs $(i,j)$ 
for which we have a
grade (i.e., student $i$ actually took course $j$).  And, of course, the
$\epsilon_{i,j}$'s are the ``errors'' one needs to add to make the approximation
an equality.
These errors reflect both the randomness associated with how any student might perform
in any particular course and also a systematic deviation between the student's
overall aptitude and his/her subject-specific aptitude for the material in the
particular course.

Ideally, grades should reflect aptitude.  Hence, we would like to say that a
student with a B-level aptitude should be expected to get B-level grades.  In
other words, inflatedness should measure deviations, both positive (for courses
with high grades) and negative (for courses with low grades), around some
neutral average grade.  In other words, we wish to impose the added constraint that
\begin{equation}
   \sum_j \nu_j = 0 .
\end{equation}
This, of course, is by choice.  We need some sort of normalization.  Without
one, we could add an arbitrary constant to every $\mu_i$ and subtract the same
constant from every $\nu_j$ without changing any of the $\epsilon_{ij}$'s.

Our aim is to find the best ``fit'' to the data.  That is, we wish to choose the
$\mu_i$'s and the $\nu_j$'s in such a manner as to make the $\epsilon_{ij}$'s as
small as possible.  To do this, we minimize the sum of the squares of
the $\epsilon_{ij}$'s:
\begin{equation}
    \begin{array}[t]{ll}
        \text{minimize }   &  
	\displaystyle \sum_{(i,j) \in \mathcal{G}} \epsilon_{ij}^2 \\[0.2in]
        \text{subject to } & 
	\begin{alignedt}
	    X_{ij} & = \mu_i + \nu_j + \epsilon_{ij} 
	          \qquad \mbox{for $(i,j) \in \mathcal{G}$}
		  \\[0.1in]
	    \displaystyle \sum_{j} \nu_j & = 0 .
        \end{alignedt}
    \end{array}
\end{equation}
Of course, we could minimize the sum of the absolute values 
instead of the sum of the squares.  
Generally speaking, sample means minimize the sum of
squares whereas sample medians minimize the sum of absolute deviations.  
Medians are more robust estimators of centrality than means but it is easier to
provide confidence intervals for means.  
For the latter reason, we will stick with
summing squares for most of this paper.


Table \ref{tab4} shows the output for Beatle University.
\begin{table}
\begin{center}
\begin{tabular}{l|cccccc|ll}
         & MAT   & CHE   & ANT   & REL   & POL   & ECO  & GPA & $\pb \mu_i$ \\[0.1in] 
  \hline
  John  & B$-$  & B$\pz$  & B$+$  & A$-$  &         &      & $3.18$ & $3.51$ \\
  Paul  & C$+$  & B$-$    &       & B$+$  & A$-$    &      & $3.00$ & $3.16$ \\
  George&       & C$+$    & B$-$  &       & B$+$    & A$-$ & $3.00$ & $2.84$ \\
  Ringo &       &         & C$+$  & B$-$  & B$\pz$  & B$+$ & $2.83$ & $2.49$ \\
  \hline
  Avg.   & $\pz2.50$ & $\pz2.70$ & $\pz2.77$& $\pz3.23$& $\pz3.33$& $\pz3.50$  \\
  $\pb \nu_j$ & $-0.84$ & $-0.50$ & $-0.18$ & $+0.18$ & $+0.50$ & $+0.84$
\end{tabular}
\end{center}
\vspace*{0.2in}
\caption{The same example as shown in Table \ref{tab1} with aptitude $\mu_i$ and
inflatedness $\nu_j$ shown alongside row and column grade averages.}
\label{tab4}
\end{table}
The student aptitude metrics clearly show that John is the smartest Beatle.
Also, while average grades in the courses correctly show that Math is the most
difficult and Econ is the easiest, the inflatedness metric expands on the
disparity.  For example, based on averages, a student might think that the
difference between Math and Econ is just one full letter grade but the
inflatedness metric suggests the difference is more like one and two thirds
letter grades ($1.68$ to be precise).

We will return to more examples later in Section \ref{sec7} including one
example using real-world data.   But, first, let us analyze our model.

\section{Least Squares}

Statistical estimates of underlying unobserved fundamental quantities have
little value without an associated estimate for an error in the estimation.  
For general least squares models, 
it is well understood how to produce such error bars.  
Nonetheless, it
is instructive to derive the formulae from scratch in this particular context, 
at least in the particular case where we assume,
unrealistically, that every student takes every course.

\subsection{Estimating Means}
To make a connection with utterly standard and elementary concepts, let us assume
for the moment that
we simply want to estimate some underlying single parameter $\mu$ based on $n$
observations $X_j$, $j=1,2,\ldots,n$.  In other words, we assume that
\[
    X_j = \mu + \epsilon_j
\]
where the $\epsilon_j$'s are taken to be independent, identically-distributed
random variables with mean zero and variance $\sigma^2$.  The parameter $\mu$ is
unknown and to be estimated.   The variance $\sigma^2$ is also unknown and must
be estimated as well.  The least squares estimator $\mubar$ for $\mu$ is that
value of $\mu$ that minimizes 
\[
    f(\mu) = \frac{1}{n} \sum_j \left( X_j - \mu \right)^2 .
\]
Taking the derivative and setting it equal to zero, one gets that $\mubar$ is
just the sample mean:
\[
    \mubar = \frac{1}{n} \sum_j X_j .
\]
Since the $X_j$'s are independent and have variance $\sigma^2$, it follows that
$\mubar$ has variance $\sigma^2/n$.
The function $f$ evaluated at $\mubar$ provides a good estimator for
$\sigma^2$:
\[
    \sigma^2 \approx f(\mubar) 
	      = \frac{1}{n} \sum_j \left( X_j - \frac{1}{n} \sum_k X_k \right)^2
	      .
\]

\subsection{Every Student Takes Every Course}

Now let's consider the problem of estimating aptitude and inflatedness from
grade data.  But, in an attempt to keep things simple, let us assume that every
student takes every course.  We have $m$ students and $n$ courses and therefore
the set $\mathcal{G}$ consists of $mn$ pairs for which we have grades.  
As before, let $f$ denote the function to be minimized:
\[
    f(\mu_1,\ldots,\mu_m,\nu_1,\ldots,\nu_n)
	= \frac{1}{mn} \sum_{i,j} \left( X_{ij} - \mu_i - \nu_j \right)^2 .
\]
As mentioned earlier, there is an ambiguity in the model---we could add an
arbitrary constant to every aptitude and subtract that same constant from every
inflatedness and the function $f$ would be unchanged.   In a previous section,
we addressed
this ambiguity by imposing one extra constraint, namely, that the sum of the
$\nu_j$'s be zero.   We could do that here, introducing then the associated
Lagrange multiplier, forming the Lagrangian, and solving the problem that way.
But, it is such a simple constraint that we prefer to introduce it in a less
formal manner as we go.  In doing so, we hope that the analysis will be more
transparent, not less.

Taking derivatives with respect to each of the variables and setting these
derivatives to zero, we get the following system of equations for the
estimators $\mubar_j$'s and $\nubar_i$'s:
\begin{eqnarray*}
    \mubar_i & = & \frac{1}{n} \sum_j \left( X_{ij} - \nubar_j \right) \\
    \nubar_j & = & \frac{1}{m} \sum_i \left( X_{ij} - \mubar_i \right) .
\end{eqnarray*}
Here, it is convenient to switch to matrix-vector notation.  So, letting
\[
    \mubar = \left[ \begin{array}{c} 
                 \mubar_1 \\ \mubar_2 \\ \vdots \\ \mubar_m 
	     \end{array} \right] ,
	     \quad 
    \nubar = \left[ \begin{array}{cccccc} 
                 \nubar_1 & \nubar_2 & \cdots & \nubar_n 
	     \end{array} \right] ,
\]
and
\[
    X = \left[ \begin{array}{cccccc} 
                 X_{11} & X_{12} & \cdots & X_{1n} \\
                 X_{21} & X_{22} & \cdots & X_{2n} \\
		 \vdots & \vdots &        & \vdots \\
                 X_{m1} & X_{m2} & \cdots & X_{mn}  \\
	     \end{array} \right] ,
\]
we can rewrite our optimality equations as
\begin{eqnarray}
    \mubar & = & \frac{1}{n} \left( Xe - e \nubar e \right) \nonumber \\
    \nubar & = & \frac{1}{m} \left( e^T X - e^T \mubar e^T \right) \label{4} ,
\end{eqnarray}
where $e$ denotes a column vector of either $m$ or $n$ ones, the dimension being
obvious from context.  
Substituting the second equation into the first, we can isolate $\mubar$:
\[
    \mubar = \frac{1}{n} \left( Xe - 
		    \frac{1}{m} e \left( e^T X - e^T \mubar e^T \right) 
		    e \right) .
\]
Collecting terms involving $\mubar$ on the left side, the remaining terms on
the right-hand side, and using the fact that $e^T e = n$, we get
\[
    \left( I - \frac{1}{m} e e^T \right) \mubar 
    = 
    		\left( 
		    I
		    - 
		    \frac{1}{m} e e^T 
		    \right) 
    		\left( \frac{1}{n} 
		    Xe  \right)
    .
\]
If the matrix $I - ee^T/m$ were nonsingular, we would at this point conclude
that 
\begin{equation} \label{1}
    \mubar = \frac{1}{n} Xe .
\end{equation}
But, the matrix is singular with rank deficiency one ($e$ is in the null space).
So, there are other choices for $\mubar$.   Indeed, there is a one-parameter
family of choices (any $\mubar$ for which $\mubar - (1/n)Xe$ is in the null
space of $I - ee^T/m$).  Nonetheless, we choose to let $\mubar$ be given by
\eqref{1} and as we shall now show this choice guarantees that the sum of the
$\nubar_j$'s vanishes as we have required.  Indeed, plugging \eqref{1} into
\eqref{4}, we get
\begin{equation} \label{6}
    \nubar = \frac{1}{m} \left( e^T X - \frac{1}{n} e^T X e e^T \right)  
\end{equation}
and therefore that
\[
    \nubar e = \frac{1}{m} \left( e^T X - \frac{1}{n} e^T X e e^T \right) e 
             = \frac{1}{m} \left( e^T X e - e^T X e \right)  = 0, 
\]
the second equality following from the fact that $e^T e = n$.

From \eqref{1} and \eqref{6}, we see that the $\mubar_i$'s and the $\nubar_j$'s
are just row and column sample means with one of them shifted by the overall
mean.

Reverting back to explicit component notation, \eqref{1} and \eqref{6} can be
written as
\begin{eqnarray*}
    \mubar_i & = & \frac{1}{n} \sum_j X_{ij}, \qquad i=1,2,\ldots,m, \\
    \nubar_j & = & \frac{1}{m} \sum_i X_{ij} - \frac{1}{mn} \sum_{i,j} X_{ij},
    							\qquad j=1,2,\ldots,n .
\end{eqnarray*}
From the first formula, we immediately see that
\begin{equation} \label{10a}
    \var(\mubar_j) = \frac{\sigma^2}{n} .
\end{equation}
Computing the variance of the $\nubar_j$'s is a little more tedious but entirely
routine.  The result is
\begin{equation} \label{11a}
    \var(\nubar_i) = \frac{\sigma^2}{m} \left( 1-\frac{1}{n} \right) 
	           \approx \frac{\sigma^2}{m} .
\end{equation}

Finally, we need an estimate of $\sigma^2$.  As before, we can use 
the objective function $f$ evaluated at the optimal values for the $\mu_i$'s and
$\nu_j$'s:
\[
    \sigma^2 
    \approx 
    f(\mubar_1,\ldots,\mubar_m,\nubar_1,\ldots,\nubar_n)
    = 
    \frac{1}{mn} \sum_{i,j} \left( X_{ij} - \mubar_i - \nubar_j \right)^2 
      .
\]

\subsection{Students Take Selected Courses}

Now suppose that each student takes only a small subset of the courses offered.
For each student $i$, let $\Ji$ denote the set of courses taken by student $i$.
Similarly, for each course $j$, let $\Ij$ denote the set of students that took
course $j$.

The least-squares loss function is now given by
\[
    f(\mu_1,\ldots,\mu_m,\nu_1,\ldots,\nu_n)
	= \frac{1}{N} \sum_{(i,j) \in \mathcal{G}} 
	           \left( X_{ij} - \mu_i - \nu_j \right)^2 ,
\]
where $N$ denotes the cardinality of the grade-set $\mathcal{G}$.
Again, we differentiate and set to zero.  This time we get
\begin{eqnarray}
    \mubar_i & = & \frac{1}{n_i} \sum_{j \in \Ji} \left( X_{ij} - \nubar_j \right) 
		\qquad i=1,2,\ldots,m \label{10} \\
    \nubar_j & = & \frac{1}{m_j} \sum_{i \in \Ij} \left( X_{ij} - \mubar_i \right) 
		\qquad j=1,2,\ldots,n \label{11} ,
\end{eqnarray}
where $n_i$ denotes the cardinality of $\Ji$ and $m_j$ denotes the cardinality
of $\Ij$.
Substituting \eqref{11} into \eqref{10}, we get
\[
    \mubar_i = \frac{1}{n_i} \sum_{j \in \Ji} \left( X_{ij} - 
		    \frac{1}{m_j} \sum_{i' \in \Ij} \left( X_{i'j} - \mubar_{i'} \right) 
		    \right) 
		\qquad i=1,2,\ldots,m .
\]
This is a set of $m$ equations in $m$ unknowns.  If there is adequate diversity
in student course selections so that every course indirectly is connected to
every other course, then one would expect this system to have rank $m-1$ leaving
only one dimensional ambiguity in the equations.  Inspired by the simplicity of
the results in the previous section, we can hope that again simple sample means
will provide one solution to this
system of equations:
\[
    \mubar_i \stackrel{\text{?}}{=} \frac{1}{n_i} \sum_{j \in \Ji} X_{ij} 
		\qquad i=1,2,\ldots,m .
\]
In order for this to be correct, we need to have
\[
    \sum_{j \in \Ji} \frac{1}{m_j} 
    \sum_{i' \in \Ij} \left( X_{i'j} - 
                    \frac{1}{n_{i'}} \sum_{j' \in \Jip} X_{i'j'} 
	    \right)
    = 0 .
\]
Unfortunately, there is no particular reason for this to be true.  And, as we
saw with the second example in the introduction, it is possible for the sample
means to be all the same even when there is a big difference in course grade
inflatedness and/or in student aptitude.  The model detects such differences.

Even though it appears there is no simple formula for the solution to the
least-squares formulation of our problem, modern statistical and/or optimization
software can solve these problems numerically
without difficulty even when the data sets are very large.

Also, the fact that we have not been able to give a simple concrete formula for
the $\mubar_i$'s and the $\nubar_j$'s makes it impossible to give a simple
concrete formula for the variance of these random variables.  Nonetheless, we
can infer from the concrete results obtained before that one should first
estimate $\sigma^2$ using the optimal value of the objective function as an
estimate of this quantity and then the variance of the individual $\mubar_i$'s
and $\nubar_j$'s can be approximated simply by dividing by the number of grades
reflected in that aggregation (that is, either $n_i$ or $m_j$).

\section{Least Absolute Deviations}

In this section, we consider a robust model in which we minimize the sum of
the absolute deviations.   To motivate what follows, we start with a brief
review of medians.

\subsection{Medians}
As when we discussed means, let us assume for the moment that
we simply want to estimate some underlying single parameter $\mu$ based on $n$
observations $X_j$, $j=1,2,\ldots,n$.  In other words, we assume that
\[
    X_j = \mu + \epsilon_j
\]
where the $\epsilon_j$'s are taken to be independent, identically-distributed
random variables with mean zero and variance $\sigma^2$.  
The least absolute deviation estimator $\muhat$ for $\mu$ is the
value of $\mu$ that minimizes 
\[
    f(\mu) = \frac{1}{n} \sum_j \left| X_j - \mu \right| .
\]
Taking the derivative and setting it equal to zero, 
one gets that $\muhat$ must satisfy
\[
    \sum_j \sgn( X_j - \muhat ) = 0,
\]
which is clearly solved by setting $\muhat$ equal to the median of the $X_j$'s (so
that half of the $\sgn$'s are $+1$ and the other half are $-1$.

\subsection{Every Student Takes Every Course}

Now let's consider the problem of estimating student aptitude and 
course inflatedness from
grade data.  As before, we start by assuming that every
student takes every course.  
Once again, let $f$ denote the function to be minimized:
\[
    f(\mu_1,\ldots,\mu_m,\nu_1,\ldots,\nu_n)
	= \frac{1}{mn} \sum_{i,j} \left| X_{ij} - \mu_i - \nu_j \right| .
\]

Taking derivatives with respect to each of the variables and setting these
derivatives to zero, we get the following system of equations for the
estimators $\muhat_j$'s and $\nuhat_i$'s:
\begin{eqnarray*}
    & \sum_j \sgn ( X_{ij} - \muhat_i - \nuhat_j ) = 0 & \qquad i=1,2,\ldots,m \\
    & \sum_i \sgn ( X_{ij} - \muhat_i - \nuhat_j ) = 0 & \qquad j=1,2,\ldots,n .
\end{eqnarray*}
Unlike before, there seems to be no simple description of the solution to this
problem.  But, we can give an algorithm that should converge quickly to the
solution.   Specifically, initialize
\begin{eqnarray*}
    \nuhat_j & = & 0, \qquad \qquad  \qquad j = 1, 2, \ldots, n \\
    \muhat_i & = & \median\{X_{ij} \;|\; j=1,2,\ldots n \}, \quad i = 1,2,\ldots,m .
\end{eqnarray*}
Then, iterate the following until there is no change from one iteration to the
next:
\begin{eqnarray*}
    \nuhat_j & = & \median\{X_{ij} - \muhat_i \;|\; i=1,2,\ldots,m\}, 
		\quad j=1,2,\ldots,n \\
    \muhat_i & = & \median\{X_{ij} - \nuhat_j \;|\; j=1,2,\ldots,n\}, 
		\quad i=1,2,\ldots,m .
\end{eqnarray*}
This algorithm is unlikely to converge to a solution that satisfies
$\sum_j \nuhat_j = 0$ but, given the initialization, it should come close to
this point.

\subsection{Students Take Selected Courses}

Finally, let us return to the general case in which 
each student takes only a small subset of the courses offered.
The problem is to minimize the sum of the absolute values of
the $\epsilon_{ij}$'s:
\begin{equation}
    \begin{array}[t]{ll}
        \text{minimize }   &  
	\displaystyle \sum_{(i,j) \in \mathcal{G}} \left|\epsilon_{ij}\right| 
						\\[0.2in]
        \text{subject to } & 
	\begin{alignedt}
	    X_{ij} & = \mu_i + \nu_j + \epsilon_{ij} 
	          \qquad \mbox{for $(i,j) \in \mathcal{G}$}
		  \\[0.1in]
	    \displaystyle \sum_{j} \nu_j & = 0 .
        \end{alignedt}
    \end{array}
\end{equation}

It is easy to reformulate this model as a linear programming (LP) problem:
\[
    \begin{array}{ll}
        \text{minimize }   &  
	\qquad \displaystyle \sum_{(i,j) \in \mathcal{G}} t_{ij} \\[0.2in]
        \text{subject to } & 
	\begin{alignedt}
	    -t_{ij} \le X_{ij} - \mu_i - \nu_j & \le t_{ij}
	          \qquad \mbox{for $(i,j) \in \mathcal{G}$}
		  \\[0.2in]
	    \displaystyle \sum_j \nu_j & = 0 .
        \end{alignedt}
    \end{array}
\]
Such linear programming problems can be solved easily.  In the next section we
give some examples and we compare the results from least squares formulations
with those from the least absolute deviations model.

\section{Examples} \label{sec7}

Finally, we consider a few specific examples including one based on real data.

\subsection{Truncated Example}

The example shown in Table \ref{tab2} was contrived in order to make a point.
In particular, it had an infinite number of students and courses.
In Table \ref{tab21}, we show a truncated version consisting of eight students
taking courses from a school offering eight courses.  Each student takes three
to five courses.   As with the untruncated version, it is clear that the
students are listed in order of their aptitude with the best student at the top.
However, student GPA's hardly reflect the obvious trend in aptitude.   The $\mu_i$'s
computed by our model make the difference in aptitude much more apparent.
Similarly, average grades given in the courses show a small trend in the
correct direction but they hardly account for the rather obvious overall trend
in course inflatedness as one scans from left to right across the table.
The $\nu_j$'s do a much better job of identifying course inflatedness.

It is interesting to point out that the least squares and the least
absolute deviation models both give the same results for this particular
example.

\begin{table}
\begin{center}
\begin{tabular}{l|cccccccc|ll}
        & MAT    & CHE    & ANT    & REL    & POL    & ECO    & HIS    & SOC    & GPA   & $\mu_i$ 
	\\[0.1in] \hline
Sean    & B$+$   & A$-$   & A$\pz$ &        &        &        &        &        & $3.67$ & $4.50$ \\
Yoko    & B$\pz$ & B$+$   & A$-$   & A$\pz$ &        &        &        &        & $3.50$ & $4.17$ \\
John    & B$-$   & B$\pz$ & B$+$   & A$-$   & A$\pz$ &        &        &        & $3.33$ & $3.83$ \\
Paul    &        & B$-$   & B$\pz$ & B$+$   & A$-$   & A$\pz$ &        &        & $3.33$ & $3.50$ \\
George  &        &        & B$-$   & B$\pz$ & B$+$   & A$-$   & A$\pz$ &        & $3.33$ & $3.17$ \\
Ringo   &        &        &        & B$-$   & B$\pz$ & B$+$   & A$-$   & A$\pz$ & $3.33$ & $2.83$ \\
Jane    &        &        &        &        & B$-$   & B$\pz$ & B$+$   & A$-$   & $3.17$ & $2.50$ \\
Heather &        &        &        &        &        & B$-$   & B$\pz$ & B$+$   & $3.00$ & $2.17$ \\
	\hline
Avg. & $\pz3.00$ & $\pz3.17$ & $\pz3.33$ & $\pz3.33$ & $\pz3.33$ & $\pz3.33$ & $\pz3.50$ & $\pz3.67$ \\
$\pb \nu_j$ & $-1.17$ & $-0.83$ & $-0.50$ & $-0.17$ & $+0.17$ & $+0.50$ & $+0.83$ & $+1.17$ \\
\end{tabular}
\end{center}
\vspace*{0.2in}
\caption{{\em Truncated Example.} 
	This is the same as the example shown in Table \ref{tab2} but it has
	been truncated to represent a school with eight students and eight courses.  
	Each student took three to five courses with grades as shown.
	As with the untruncated version, there are clear trends in student
	aptitude and course inflatedness, which our model correctly uncovers.
}
\label{tab21}
\end{table}

\subsection{Circulant Example}

This example is almost the same as the truncated example in the previous
subsection.   Here, however, we have added two courses to Sean's schedule and to
Heather's schedule and we have added one course to Yoko's schedule and to Jane's
schedule.  The result is a table of grades that has a {\em circulant} structure.
Now, the trends that were clearly apparent in the truncated example are
completely gone.   In this example, both student GPA and the $\mu_i$'s reflect
the lack of any differentiation among the students.   Similarly, the course
averages and the $\nu_j$'s both show that all courses are curved the same.

\begin{table}
\begin{center}
\begin{tabular}{l|cccccccc|ll}
        & MAT    & CHE    & ANT    & REL    & POL    & ECO    & HIS    & SOC    & GPA   & $\mu_i$ 
	\\[0.1in] \hline
Sean    & B$+$   & A$-$   & A$\pz$ &        &        &        & B$-$   & B$\pz$ & $3.33$ & $3.33$ \\
Yoko    & B$\pz$ & B$+$   & A$-$   & A$\pz$ &        &        &        & B$-$   & $3.33$ & $3.33$ \\
John    & B$-$   & B$\pz$ & B$+$   & A$-$   & A$\pz$ &        &        &        & $3.33$ & $3.33$ \\
Paul    &        & B$-$   & B$\pz$ & B$+$   & A$-$   & A$\pz$ &        &        & $3.33$ & $3.33$ \\
George  &        &        & B$-$   & B$\pz$ & B$+$   & A$-$   & A$\pz$ &        & $3.33$ & $3.33$ \\
Ringo   &        &        &        & B$-$   & B$\pz$ & B$+$   & A$-$   & A$\pz$ & $3.33$ & $3.33$ \\
Jane    & A$\pz$ &        &        &        & B$-$   & B$\pz$ & B$+$   & A$-$   & $3.33$ & $3.33$ \\
Heather & A$-$   & A$\pz$ &        &        &        & B$-$   & B$\pz$ & B$+$   & $3.33$ & $3.33$ \\
	\hline
Avg. & $3.33$ & $3.33$ & $3.33$ & $3.33$ & $3.33$ & $3.33$ & $3.33$ & $3.33$ \\
$\pb \nu_j$ & $0.00$ & $0.00$ & $0.00$ & $0.00$ & $0.00$ & $0.00$ & $0.00$ & $0.00$ \\
\end{tabular}
\end{center}
\vspace*{0.2in}
\caption{{\em Circulant Example.} 
	This example is the same as the previous one except that there are six
	more grades filling out the matrix into a circulant form.
	Now the trends are gone.
	Every student has a B+ average and every course is curved to a B+.  Our
	model correctly assigns every course an easiness adjustment of $0.00$
	leaving every student's ``corrected'' GPA equal to his/her original GPA.
}
\label{tab22}
\end{table}

\include{MC}

\subsection{Two Semesters of Real Data}

The registrar at a private university in the northeast has given us a complete
two-semester data set.
There are about 5000 students at this university each of
whom takes four or five courses per semester from a selection of roughly 700 courses offered
each semester.
The data is encoded---we don't know the identity of any particular student.  Nor
can we tell which course is which.   All of this has been pre-encoded by the
registrar.   But, the grades are real.
A small snippet of the data is shown in Table \ref{tab3}.

Table \ref{tab5} shows
a sample of the output from the least squares model.
Table \ref{tab10} shows
a sample of the output from the least absolute deviations model.
Comparing Tables \ref{tab5} and \ref{tab10}, it is clear that the results are
similar.
\begin{table}
\begin{center}
\small
\begin{minipage}{1.6in}
\begin{tabular}{rll|}
1 & F001090 & 3.7 \\ 
1 & F004148 & 1.7 \\ 
1 & F006665 & 2 \\ 
1 & F010449 & 3 \\ 
1 & S009167 & 3.7 \\ 
1 & S009571 & 2 \\ 
1 & S010994 & 2.7 \\ 
2 & F003387 & 3 \\ 
2 & F009193 & 2.7 \\ 
2 & F010693 & 3 \\ 
2 & F010813 & 2.7 \\ 
2 & S001093 & 1 \\ 
2 & S003408 & 3 \\ 
2 & S005302 & 2 \\ 
3 & F003769 & 3 \\ 
3 & F004893 & 3.7 \\ 
3 & F004896 & 3.3 \\ 
3 & S004172 & 1.7 \\ 
4 & F000613 & 3.7 \\ 
4 & F001381 & 2.7 \\ 
4 & F004140 & 3 \\ 
4 & F005588 & 3.7 \\ 
4 & S000185 & 3 \\ 
4 & S004398 & 3 \\ 
4 & S004901 & 2.7 \\ 
4 & S009698 & 3.3 \\
\end{tabular}
\end{minipage}
\begin{minipage}{1.6in}
\begin{tabular}{lll|}
5 & F002046 & 3 \\ 
5 & F005976 & 3.7 \\ 
5 & F007285 & 2.3 \\ 
5 & F008991 & 4 \\ 
5 & F010762 & 4 \\ 
5 & S001380 & 3.7 \\ 
5 & S004153 & 2.3 \\ 
5 & S005842 & 4 \\ 
5 & S008310 & 4 \\ 
6 & F001400 & 2.7 \\ 
6 & F004647 & 3.7 \\ 
6 & F006787 & 3.3 \\ 
6 & F009999 & 2.7 \\ 
6 & S003424 & 3 \\ 
6 & S003952 & 3 \\ 
6 & S009187 & 3.7 \\ 
6 & S010953 & 3.7 \\ 
7 & F005979 & 3.3 \\ 
7 & F007230 & 3.3 \\ 
7 & F010437 & 3.3 \\ 
7 & S006804 & 4 \\ 
7 & S010960 & 4 \\ 
8 & F001064 & 4 \\ 
8 & F002461 & 2.3 \\ 
8 & F005979 & 3 \\ 
8 & S007946 & 0 \\ 
\end{tabular}
\end{minipage}
\begin{minipage}{1.6in}
\begin{tabular}{lll}
8 & S008811 & 2.7 \\ 
8 & S010952 & 3.7 \\ 
8 & S010973 & 1.7 \\ 
9 & F002614 & 2.7 \\ 
9 & F006664 & 2 \\ 
9 & F008144 & 1.7 \\ 
9 & F008832 & 2.7 \\ 
9 & F010542 & 3 \\ 
9 & S001065 & 3.3 \\ 
9 & S001542 & 2 \\ 
9 & S004398 & 2.3 \\ 
9 & S004399 & 2.3 \\ 
10 & F008991 & 2 \\ 
10 & F009582 & 2.3 \\ 
10 & S001404 & 4 \\ 
10 & S002463 & 4 \\ 
10 & S004186 & 4 \\ 
10 & S004398 & 2 \\ 
11 & F001090 & 3.3 \\ 
11 & F001109 & 3.7 \\ 
11 & F003243 & 1.7 \\ 
11 & F005558 & 2.3 \\ 
11 & S002625 & 2.3 \\ 
11 & S007854 & 2.7 \\ 
11 & S010979 & 3.3 \\ 
   & $\vdots$ &       \\
\end{tabular}
\end{minipage}
\end{center}
\vspace*{0.2in}
\caption{One semester of data consisting of about $37000$ grades given to about
$5000$ students.
Each record consists of three data elements: the student id (encoded), the
course id (also encoded), and the grade (converted from a letter grade to a numerical
grade in the usual manner).}
\label{tab3}
\end{table}

Typical courses have between $10$ and $100$ students.   For the larger courses, 
there seems to be an adequate amount of data to draw conclusions.
Since, the data set only represents two semesters and most students take only
four or five courses in a semester, one should not put too much credence in the
aptitudes assigned to the students.  But, a larger data set consisting of three
or four years of data would contain about $20$ to $30$ courses of grade data for
each student.  In such a case, one could imagine that the $\mu_i$'s would be a
pretty good indicator of student aptitude.

\section{Conclusions}
The fundamental data available to a registrar is grading data: the $X_{ij}$'s.
In recent years, this data set has been used for two main purposes: (1) to assess student
achievement, and (2) to assess course-by-course grade inflation.
Student achievement is usually assessed by reporting on a transcript the
student's GPA.  A statistical justification for this is that the totality of all
student GPA's is the simple least-squares solution to the
following regression model:
\[
    X_{ij} = \mu_i + \epsilon_{ij}, \qquad (i,j) \in \mathcal{G} .
\]
At the same time, grade inflation is assessed by reporting average (or median)
grades given in a course.   The totality of average course grades is the
least-squares solution to a ``dual'' regression model:
\[
    X_{ij} = \nu_j + \epsilon_{ij}, \qquad (i,j) \in \mathcal{G} .
\]
It seems only natural that these two problems should be combined into one and
that is exactly what we have proposed in this paper.

Grade inflation, and what to do about it, has been discussed extensively in
recent years.  In this paper, we have described an analytical approach to
disentangling the course-by-course differences in grading policies from
underlying student aptitudes.  If such a tool were to be widely adopted and
student aptitude as defined by the models given in this paper were to become the
accepted measure of student accomplishment, then the issue of standardizing
grading policies across a university becomes somewhat moot.   

Of course, there is still the important 
question of comparing grades from students across different universities,
which is something professional schools, graduate schools, and employers
must do routinely.  Unfortunately, the model described here cannot address 
this difficult problem without a dataset in which students at divergent
universities take common courses.  Perhaps the only way to do that would be to
make a huge model in which all high-school and university grading data are fed
into one huge master program.   If such data were ever made available, which is
highly doubtful, such a problem would probably be too large to solve on today's
computers.

The models presented in this paper are good examples of least-squares and
least-absolute deviations regression and can therefore be used as a pedagogical
tool when teaching these topics in statistics and/or optimization courses.

\section{Further Reading}

There is, of course, prior literature on the general problem of assessment.
Rasch's book \cite{Rasch60} and the related paper \cite{Rasch61} introduce, perhaps
for the first time, the idea of representing a score as a function of the difference between
ability and difficulty.
Caulkins et al. \cite{CLW96} apply the idea specifically to the problem of adjusting
grade-point averages.  Johnson \cite{Joh97} introduced an alternative approach and
compared it to the linear-adjustment models.  More recently, the book \cite{Joh03} 
gives an extensive treatment on a number of models for adjusting for variations
in course difficulty.

\begin{table}
\begin{center}
\begin{minipage}{5.4in}
\begin{tabular}{lcr|lcr|lcr}
F001204 & $ -2.55 \pm  0.50 $ & $    1 $  &	S008128 & $ -0.54 \pm  0.36 $ & $    2 $ &	F010864 & $  0.64 \pm  0.29 $ & $    3 $ \\
F002509 & $ -2.49 \pm  0.50 $ & $    1 $  &	F002339 & $ -0.53 \pm  0.11 $ & $   22 $ &	S002603 & $  0.64 \pm  0.13 $ & $   14 $ \\
S001225 & $ -1.77 \pm  0.50 $ & $    1 $  &	F004137 & $ -0.53 \pm  0.10 $ & $   25 $ &	S008485 & $  0.64 \pm  0.12 $ & $   17 $ \\
S003935 & $ -1.04 \pm  0.36 $ & $    2 $  &	F008314 & $ -0.53 \pm  0.15 $ & $   11 $ &	F000295 & $  0.65 \pm  0.13 $ & $   14 $ \\
F003936 & $ -0.89 \pm  0.50 $ & $    1 $  &	F009959 & $ -0.53 \pm  0.13 $ & $   14 $ &	F010480 & $  0.65 \pm  0.25 $ & $    4 $ \\
S002963 & $ -0.86 \pm  0.09 $ & $   33 $  &	F010275 & $ -0.52 \pm  0.29 $ & $    3 $ &	S010396 & $  0.66 \pm  0.29 $ & $    3 $ \\
S005818 & $ -0.77 \pm  0.17 $ & $    9 $  &	S008328 & $ -0.52 \pm  0.15 $ & $   11 $ &	F010501 & $  0.68 \pm  0.50 $ & $    1 $ \\
S004319 & $ -0.75 \pm  0.23 $ & $    5 $  &	F005558 & $ -0.51 \pm  0.04 $ & $  187 $ &	F002968 & $  0.69 \pm  0.50 $ & $    1 $ \\
S008329 & $ -0.70 \pm  0.15 $ & $   12 $  &	S000519 & $ -0.51 \pm  0.15 $ & $   12 $ &	F009955 & $  0.69 \pm  0.36 $ & $    2 $ \\
S003007 & $ -0.68 \pm  0.21 $ & $    6 $  &	S001093 & $ -0.50 \pm  0.07 $ & $   47 $ &	S007268 & $  0.69 \pm  0.17 $ & $    9 $ \\
S001783 & $ -0.66 \pm  0.08 $ & $   36 $  &	S008624 & $ -0.50 \pm  0.19 $ & $    7 $ &	S010988 & $  0.73 \pm  0.18 $ & $    8 $ \\
S010294 & $ -0.66 \pm  0.29 $ & $    3 $  &	F001003 & $ -0.49 \pm  0.19 $ & $    7 $ &	F010535 & $  0.74 \pm  0.50 $ & $    1 $ \\
F004151 & $ -0.65 \pm  0.05 $ & $  107 $  &	F002060 & $ -0.49 \pm  0.12 $ & $   18 $ &	F010783 & $  0.75 \pm  0.36 $ & $    2 $ \\
F008345 & $ -0.60 \pm  0.15 $ & $   11 $  &	$\cdots$ & $\cdots$           &    	 &	S008506 & $  0.78 \pm  0.29 $ & $    3 $ \\
S002477 & $ -0.60 \pm  0.17 $ & $    9 $  &	S001543 & $  0.58 \pm  0.16 $ & $   10 $ &	S009990 & $  0.78 \pm  0.29 $ & $    3 $ \\
S004159 & $ -0.60 \pm  0.08 $ & $   42 $  &	S007263 & $  0.58 \pm  0.23 $ & $    5 $ &	S010720 & $  0.78 \pm  0.29 $ & $    3 $ \\
F004140 & $ -0.59 \pm  0.05 $ & $  122 $  &	S010725 & $  0.58 \pm  0.19 $ & $    7 $ &	S010987 & $  0.80 \pm  0.18 $ & $    8 $ \\
F008328 & $ -0.59 \pm  0.16 $ & $   10 $  &	S010932 & $  0.58 \pm  0.29 $ & $    3 $ &	F010617 & $  0.81 \pm  0.21 $ & $    6 $ \\
S001380 & $ -0.59 \pm  0.03 $ & $  312 $  &	F010402 & $  0.59 \pm  0.18 $ & $    8 $ &	F010830 & $  0.84 \pm  0.29 $ & $    3 $ \\
F004153 & $ -0.58 \pm  0.15 $ & $   11 $  &	S004063 & $  0.60 \pm  0.23 $ & $    5 $ &	S010986 & $  0.90 \pm  0.50 $ & $    1 $ \\
S009395 & $ -0.58 \pm  0.29 $ & $    3 $  &	S004870 & $  0.60 \pm  0.25 $ & $    4 $ &	S000205 & $  0.93 \pm  0.50 $ & $    1 $ \\
F009200 & $ -0.57 \pm  0.18 $ & $    8 $  &	F005922 & $  0.61 \pm  0.21 $ & $    6 $ &	S011047 & $  0.96 \pm  0.29 $ & $    3 $ \\
F010277 & $ -0.57 \pm  0.25 $ & $    4 $  &	F010395 & $  0.61 \pm  0.29 $ & $    3 $ &	S010039 & $  1.06 \pm  0.50 $ & $    1 $ \\
F004322 & $ -0.56 \pm  0.07 $ & $   55 $  &	F004189 & $  0.62 \pm  0.25 $ & $    4 $ &	F003038 & $  1.22 \pm  0.50 $ & $    1 $ \\
F005128 & $ -0.55 \pm  0.03 $ & $  256 $  &	S001263 & $  0.62 \pm  0.18 $ & $    8 $ &	S010261 & $  1.66 \pm  0.36 $ & $    2 $ \\
S004150 & $ -0.55 \pm  0.03 $ & $  217 $  &	F004043 & $  0.63 \pm  0.21 $ & $    6 $ &	F010122 & $  1.92 \pm  0.50 $ & $    1 $ \\
\end{tabular}
\end{minipage}
\end{center}
\vspace*{0.2in}
\caption{A partial listing of the course inflatedness associated with the data
partially shown in Table \ref{tab3}.  The table shows in three columns the
beginning and the end of a long table of data with three columns.
The first column is the course id, the second column is the inflatedness
$\nu_j$, and the third column shows the course enrollment.
In the interest of space, we show only some of the least inflated
courses and some of the most inflated courses.  It is interesting to note that,
with the exception of a few very small classes (seminar and project courses), 
the inflatedness spans from about $-0.45$ to $0.55$.
In other words, a student can expect a plus/minus half-letter grade
deviation from his/her ``true'' aptitude simply because of differences
in grading policies among some courses.  
}
\label{tab5}
\end{table}

\begin{table}
\begin{center}
\begin{minipage}{5.4in}
\begin{tabular}{lcr|lcr|lcr}
F001204 & $ -3.22 \pm  0.36 $ & $    1 $ &	F001392 & $ -0.46 \pm  0.06 $ & $   38 $ &	S004206 & $  0.64 \pm  0.18 $ & $    4 $ \\
F002509 & $ -2.49 \pm  0.36 $ & $    1 $ &	F001403 & $ -0.46 \pm  0.03 $ & $  171 $ &	S005917 & $  0.64 \pm  0.11 $ & $   10 $ \\
S001225 & $ -1.78 \pm  0.36 $ & $    1 $ &	F001759 & $ -0.46 \pm  0.06 $ & $   37 $ &	F005099 & $  0.67 \pm  0.15 $ & $    6 $ \\
S003935 & $ -1.30 \pm  0.25 $ & $    2 $ &	F002376 & $ -0.46 \pm  0.25 $ & $    2 $ &	S003046 & $  0.67 \pm  0.18 $ & $    4 $ \\
F010315 & $ -1.06 \pm  0.21 $ & $    3 $ &	F002969 & $ -0.46 \pm  0.08 $ & $   19 $ &	S010342 & $  0.70 \pm  0.25 $ & $    2 $ \\
F003936 & $ -0.87 \pm  0.36 $ & $    1 $ &	F004140 & $ -0.46 \pm  0.03 $ & $  122 $ &	F004043 & $  0.71 \pm  0.15 $ & $    6 $ \\
S002491 & $ -0.82 \pm  0.21 $ & $    3 $ &	F004148 & $ -0.46 \pm  0.06 $ & $   40 $ &	F000295 & $  0.74 \pm  0.10 $ & $   14 $ \\
S002963 & $ -0.76 \pm  0.06 $ & $   33 $ &	F004149 & $ -0.46 \pm  0.03 $ & $  188 $ &	F004189 & $  0.74 \pm  0.18 $ & $    4 $ \\
S008128 & $ -0.70 \pm  0.25 $ & $    2 $ &	F004150 & $ -0.46 \pm  0.04 $ & $   96 $ &	F010783 & $  0.74 \pm  0.25 $ & $    2 $ \\
F004151 & $ -0.66 \pm  0.03 $ & $  107 $ &	F004408 & $ -0.46 \pm  0.08 $ & $   18 $ &	S007263 & $  0.74 \pm  0.16 $ & $    5 $ \\
F008328 & $ -0.66 \pm  0.11 $ & $   10 $ &	F005128 & $ -0.46 \pm  0.02 $ & $  256 $ &	S009571 & $  0.74 \pm  0.21 $ & $    3 $ \\
F010275 & $ -0.66 \pm  0.21 $ & $    3 $ &	F005558 & $ -0.46 \pm  0.03 $ & $  187 $ &	F010830 & $  0.79 \pm  0.21 $ & $    3 $ \\
S005818 & $ -0.66 \pm  0.12 $ & $    9 $ &	F006660 & $ -0.46 \pm  0.05 $ & $   45 $ &	S010986 & $  0.80 \pm  0.36 $ & $    1 $ \\
F010277 & $ -0.60 \pm  0.18 $ & $    4 $ &	$\cdots$ & $\cdots$           &    	 &	F010535 & $  0.83 \pm  0.36 $ & $    1 $ \\
F009049 & $ -0.59 \pm  0.11 $ & $   11 $ &	S010987 & $  0.62 \pm  0.13 $ & $    8 $ &	F001267 & $  0.84 \pm  0.36 $ & $    1 $ \\
F009959 & $ -0.58 \pm  0.10 $ & $   14 $ &	F008752 & $  0.63 \pm  0.21 $ & $    3 $ &	F003038 & $  0.84 \pm  0.36 $ & $    1 $ \\
F001003 & $ -0.56 \pm  0.14 $ & $    7 $ &	F002968 & $  0.64 \pm  0.36 $ & $    1 $ &	F008385 & $  0.84 \pm  0.14 $ & $    7 $ \\
F004322 & $ -0.56 \pm  0.05 $ & $   55 $ &	F004923 & $  0.64 \pm  0.14 $ & $    7 $ &	S000205 & $  0.84 \pm  0.36 $ & $    1 $ \\
S001783 & $ -0.56 \pm  0.06 $ & $   36 $ &	F010501 & $  0.64 \pm  0.36 $ & $    1 $ &	S004870 & $  0.84 \pm  0.18 $ & $    4 $ \\
S004159 & $ -0.56 \pm  0.06 $ & $   42 $ &	F010617 & $  0.64 \pm  0.15 $ & $    6 $ &	S010396 & $  0.84 \pm  0.21 $ & $    3 $ \\
S005334 & $ -0.56 \pm  0.14 $ & $    7 $ &	S000522 & $  0.64 \pm  0.21 $ & $    3 $ &	S011047 & $  0.84 \pm  0.21 $ & $    3 $ \\
S008329 & $ -0.56 \pm  0.10 $ & $   12 $ &	S001543 & $  0.64 \pm  0.11 $ & $   10 $ &	S008506 & $  0.88 \pm  0.21 $ & $    3 $ \\
S008344 & $ -0.56 \pm  0.10 $ & $   12 $ &	S001550 & $  0.64 \pm  0.11 $ & $   10 $ &	S009990 & $  0.88 \pm  0.21 $ & $    3 $ \\
F004180 & $ -0.52 \pm  0.09 $ & $   16 $ &	S002603 & $  0.64 \pm  0.10 $ & $   14 $ &	S010720 & $  0.88 \pm  0.21 $ & $    3 $ \\
F009519 & $ -0.51 \pm  0.15 $ & $    6 $ &	S003304 & $  0.64 \pm  0.16 $ & $    5 $ &	S010924 & $  0.94 \pm  0.16 $ & $    5 $ \\
S008328 & $ -0.51 \pm  0.11 $ & $   11 $ &	S004063 & $  0.64 \pm  0.16 $ & $    5 $ &	S010261 & $  1.08 \pm  0.25 $ & $    2 $ \\
\end{tabular}
\end{minipage}
\end{center}
\vspace*{0.2in}
\caption{A partial listing of the course inflatedness associated with the data
partially shown in Table \ref{tab3} as computed using the least absolute
deviations model.}
\label{tab10}
\end{table}

\noindent

\subsection*{Acknowledgement}  
The authors would like to thank 
Jianqing Fan for useful discussions regarding 
underlying statistical ideas.

\vfill
\pagebreak
\bibliographystyle{siam}
\bibliography{../lib/refs}


\end{document}